\documentclass[twocolumn,showpacs,preprintnumbers,amsmath,amssymb,prl]{revtex4}

\usepackage[dvips]{graphicx}
\usepackage{dcolumn}
\usepackage{bm}

\begin{document}

\title{Anisotropy of in-plane magnetization due to nodal gap structure
in the vortex state}
\author{Hiroaki Kusunose}
\affiliation{Department of Physics, Tohoku University, Sendai 980-8578, Japan}

\date{\today}

\begin{abstract}
We examine the interplay between anisotropy of the in-plane magnetization and the nodal gap structure on the basis of the approximate analytic solution in the quasiclassical formalism.
We show that a four-fold oscillation appears in the magnetization, and its amplitude changes sign at an intermediate field.
The high-field oscillation originates from the anisotropy of the upper critical field, while the low-field behavior can be understood by the thermally activated quasiparticles near nodes depending on the applied field angles.
The temperature dependence of the magnetization also shows a similar sign change.
The anisotropy of the magnetization offers a possible measurement to identify the gap structure directly for a wide class of type II superconductors. 
\end{abstract}
\pacs{74.25.Op,74.25.Ha,74.25.Bt}

\maketitle

In the last two decades, a family of unconventional superconductors has extensively grown up, and it has created more demand for a direct observation of the gap structure.
Although thermodynamic quantities at low temperature follow power-law behaviors due to the presence of gap nodes\cite{Leggett75,Sigrist91}, this in principle cannot determine an absolute direction of nodes, which has a significant importance to selectively identify the pairing mechanism.
Recently, a powerful method has been developed to measure the gap nodes directly.
Namely, field angle dependences in oriented magnetic fields have shown oscillatory behaviors in the thermal conductivity\cite{Izawa01a,Izawa01b,Tanatar01a,Tanatar01b,Izawa02a,Izawa02b} and the specific heat\cite{Deguchi03,Aoki03,Park03a,Park03b,Park03c,Deguchi04}, reflecting low-energy excitations inherent from the gap structure.

An analysis of those two experiments, however, is somewhat ambiguous, and leads opposite conclusions at worst\cite{Izawa01a,Aoki03}.
This is because a dominant source of the oscillation differs depending on what fields and temperatures we consider.
For instance, at low temperature the activated quasiparticle (QP) due to the supercurrent energy shift roughly determines the oscillatory behaviors in the specific heat\cite{Vekhter99a,Won01,Kusunose04}.
On the other hand, a transport lifetime in the thermal conductivity plays an important role in intermediate temperature range, which could reverse the oscillation caused by the density of states (DOS)\cite{Yu95,Maki00,Vekhter99}.
Therefore, more such angle-resolved bulk measurement is highly required to draw a definite conclusion with a consistency among different probes. 

In this Letter, we examine the interplay between anisotropy of the in-plane magnetization and  the nodal gap structure in the whole region of the $H$-$T$ diagram, and discuss whether the magnetization is a possible probe to identify the gap structure or not.
Experimentally, the basal plane anisotropy of magnetization, $M(\phi_h)$, has already been observed for example in non-magnetic borocarbides, RNi$_2$B$_2$C (R=Lu, Y)\cite{Civale99,Kogan99}.
A remarkable fact is that the amplitude of the oscillation in $M$ changes sign from high to low fields\cite{Civale99}. Temperature dependence also exhibits a similar sign change\cite{Kogan99}.
It is worth noting that the low-$H$ and low-$T$ signs are opposite to what is expected from the known angular dependence of the upper critical field, $H_{c2}$.
Indeed, such a sign change has a close resemblance to that observed in the specific heat of Sr$_2$RuO$_4$\cite{Deguchi03,Deguchi04}, where the sign change can be ascribed to the modulation of the gap amplitude\cite{Kusunose04}.
Hence it should be squarely addressed.

In the tetragonal symmetry the standard London theory does not introduce any in-plane anisotropy to the mass tensor.
Kogan {\it et al.} argued that the nonlocal correction to the London theory gives rise to anisotropy in the mass tensor, and qualitatively reproduced the overall observed behaviors\cite{Kogan99}.
Nevertheless, it is natural to address that the nodal gap structure alone could create the anisotropic behavior in $M$.
Theoretically, the validity of the nonlocal London theory is also questionable deep in the vortex state, where an overlap of the cores plays a dominant role.
In the high-field end the anisotropy of $H_{c2}$ predominates over any other contributions.

Our approach is based on the approximate analytic solution of the quasiclassical equations near $H_{c2}$\cite{Brandt67,Pesch75}.
In this approximation, we replace the spatial dependence of the magnetic induction, ${\bm B}({\bm R})$ and the diagonal quasiclassical propagator, $g_{\bm k}({\bm R})=-i\int d\xi G/\pi$ by their average, ${\bm B}$ and $g_{\bm k}$, and the Abrikosov solution, $\Delta({\bm k},{\bm R})$ is used for the vortex lattice structure.
We consider the two-dimensional cylindrical isotropic Fermi surface for simplicity, and focus on the $d_{x^2-y^2}$-wave, i.e., $\Delta({\bm k})=\Delta\varphi({\bm k})$ with $\varphi({\bm k})=\sqrt{2}\cos(2\phi)$, where $\phi$ is the azimuthal angle of ${\bm k}$ measured from the $x$-axis.
Then, we have the analytic solution of the propagator as
\begin{equation}
g=\left[1+\frac{\sqrt{\pi}}{i}\left(\frac{u_n\Delta({\bm k})}{\omega_n}\right)W'(iu_n)\right]
^{-1/2},
\end{equation}
where $u_n=2\omega_n/\tilde{v}_\perp({\bm k})\sqrt{2|e|B}$ with the fermionic Matsubara frequency and $W(z)=e^{-z^2}{\rm erfc}(-iz)$ is the Faddeeva function.
Here $\tilde{v}_\perp$ is the component of the Fermi velocity perpendicular to the field, and
$\tilde{v}_\perp^2\propto1+2\sin^2(\phi-\phi_h-\pi/4)$ for the in-plane field measured from the nodal direction.
The prefactor of the velocity characterizing the quasi two-dimensionality can be absorbed into the definition of $H_{c2}$.
Note that only the spatial average of the gap, $\Delta({\bm k})=\overline{\Delta({\bm k},{\bm R})}$ appears in the expression.
The solution recovers the correct BCS result in the limit of $B\to0$.

Although the approximation seems to be valid near $H_{c2}$, the present author has shown that the analytic solution captures an essential physics in low-$H$ and low-$T$ regions as well if we minimize the free energy with respect to ${\bm B}$ and $\Delta$\cite{Vekhter99,Kusunose04,Kusunose04a}.
In the clean limit, the free energy measured from the normal state is given by\cite{Kusunose04a}
\begin{multline}
\Omega_{\rm SN}=\frac{(B-H)^2}{8\pi}+N_0\biggl[|\Delta|^2\ln\left(4e^\gamma n_c\right)\\
-2\pi T\sum_{n=0}^{n_cT_c/T}\langle I({\bm k},\omega_n)\rangle\biggr],
\end{multline}
with
\begin{equation}
I({\bm k},\omega_n)=\frac{2g}{1+g}\frac{u_n}{\omega_n}\sqrt{\pi}|\Delta({\bm k})|^2W(iu_n),
\end{equation}
where $N_0$ is the DOS in the normal state, $\gamma\simeq0.5772$ is the Euler's constant and $\langle\cdots\rangle$ denotes the average over the Fermi surface.
The cut-off frequency $n_c=\omega_c/2\pi T=60$ is used in the present paper.
Note that the dimensionless quantity, $H_{c2}^2/N_0T_c^2$, is proportional to $(\lambda/\xi)^2$, so that we define it as our Ginzburg-Landau (GL) parameter $\kappa^2$.

\begin{figure}
\includegraphics[width=8cm]{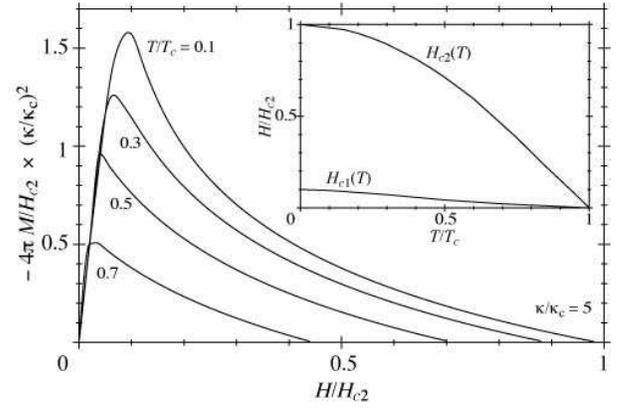}
\caption{The magnetization curves for several temperatures with $\kappa/\kappa_c=5$, where $\kappa_c$ is the critical value of the GL parameter $\kappa$ between type I and II superconductors. The inset shows $H$-$T$ diagram, where $H_{c1}(T)$ is determined by the field  showing the maximum in the magnetization curve.
}
\label{fig1}
\end{figure}

Figure~\ref{fig1} shows the magnetization curves for the field parallel to the nodal direction, i.e., $\phi_h=0$.
The GL parameter is $\kappa/\kappa_c=5$, where $\kappa_c\simeq5.363$ is the critical value between type I and II superconductors.
The calculated curves well describe the correct $H$-linear dependence near $H_{c2}$ and the logarithmic increase at low fields.
A smooth peak rather than the cusp-like behavior near $H_{c1}$ is obtained due to the artifact of the approximation.
The inset shows the $H$-$T$ phase diagram, where the $H_{c1}$ is determined by the field showing the maximum in the magnetization curve.
The Meissner state is realized below $H_{c1}$, namely, the free energy has a minimum at $B=0$.
The theory well describes the low-field behavior at low temperatures, which is the most severe region for the present approximation.

\begin{figure}
\includegraphics[width=8cm]{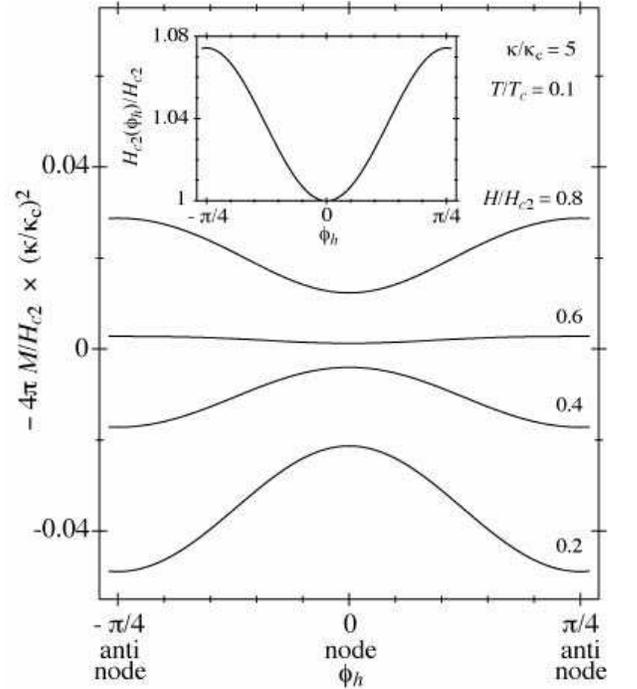}
\caption{The magnetization as a function of the field angle measured from the nodal direction. The data at different $H$ are vertically shifted.
The inset shows oscillation in $H_{c2}(\phi_h)$.}
\label{fig2}
\end{figure}

Next we discuss the field-angle dependence of the magnetization.
Figure~\ref{fig2} shows $-M$ as a function of $\phi_h$ for several $H$ at $T/T_c=0.1$.
It is remarkable that the nodal gap alone gives rise to the four-fold oscillation of $-M(\phi_h)$, which is reversed as $H$ decreases.
Note that the data at different $H$ are vertically shifted to fit in a single figure.
The inset shows the anisotropy of $H_{c2}(\phi_h)$, which is consistent with the known results.
In the present approximation at $T=0$ and $H\lesssim H_{c2}$, we have
\begin{equation}
H_{c2}(\phi_h)\propto \exp\left[-\left\langle
|\varphi({\bm k})|^2\ln\left(\tilde{v}_\perp^2({\bm k})\right)\right\rangle\right],
\end{equation}
and
\begin{equation}
-M(\phi_h)\propto \frac{1}{A(\phi_h)\kappa^2}\left(1-\frac{H}{H_{c2}(\phi_h)}\right),
\end{equation}
where $A(\phi_h)=\langle|\varphi({\bm k})|^4/\tilde{v}_\perp^2({\bm k})\rangle>0$.
From these expressions, we expect the similar oscillatory behaviors between $-M(\phi_h)$ and $H_{c2}(\phi_h)$ at high fields.

\begin{figure}
\includegraphics[width=8cm]{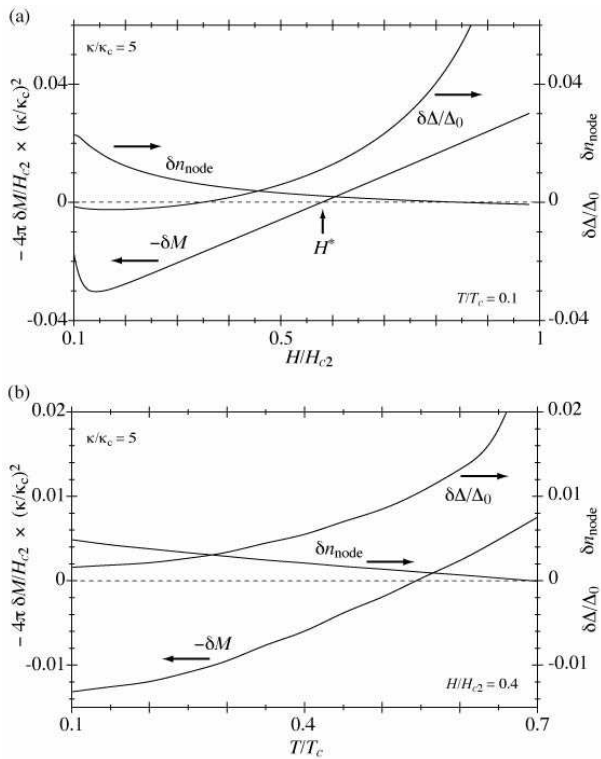}
\caption{The amplitude $-\delta M=-[M(\phi_h=\pi/4)-M(\phi_h=0)]$ of the four-fold oscillations of the in-plane magnetization as a function of (a) $H$ at $T/T_c=0.1$ and (b) $T$ at $H/H_{c2}=0.4$.
The amplitudes of the spatially averaged gap magnitude, $\delta\Delta$, and the number of the activated quasiparticles near nodes (see text in detail), $\delta n_{\rm node}$, are also shown.}
\label{fig3}
\end{figure}
To quantify the amplitude of oscillations, we define the amplitude as $-\delta M=-[M(\phi_h=\pi/4)-M(\phi_h=0)]$.
The field dependence of $-\delta M$ for $T/T_c=0.1$ is shown in Fig.~\ref{fig3}(a).
The amplitude $-\delta M$ decreases almost linearly as $H$ decreases and changes its sign at $H=H^*$.
It has already shown that the $H_{c2}$ anisotropy brings about the positive amplitude in high fields.
The sign change in decreasing field indicates that different physics comes into play to predominate over the effect of the $H_{c2}$ anisotropy.

To elucidate this point, the $H$ dependence of the oscillatory amplitude of the gap magnitude, $\delta\Delta$ and the population of the thermally activated QPs near nodes, $\delta n_{\rm node}$ are shown in Fig.~\ref{fig3}(a).
The amplitudes are defined in the same manner as $-\delta M$.
We have defined $n_{\rm node}$ normalized by the normal-state counterpart as
\begin{equation}
n_{\rm node}(\phi_h)=\sum_i^{\rm nodes} \frac{\int_0^\infty d\omega f(\omega/T) N(\omega,{\bm k}_i\pm\delta)}{4N_0T\ln2},
\end{equation}
where $N(\omega,{\bm k})={\rm Re}\;g$ is the DOS in the superconducting state, $f(x)$ is the Fermi-Dirac distribution function, and ${\bm k}_i=(\pi/2)i+\pi/4$ ($i=0,1,2,3$) with $\delta=0.02\pi$.
Note that the field-angle dependence is implicitly included in $N(\omega,{\bm k})$.
Since the gap amplitude has a strong $H$ dependence near $H_{c2}$, i.e., $\Delta(\phi_h)\propto\sqrt{1-H/H_{c2}(\phi_h)}$, the anisotropy in $H_{c2}(\phi_h)$ yields the larger anisotropy in $\Delta(\phi_h)$ and hence the larger $-M(\phi_h)$ at closer $H_{c2}$ as shown in Fig.~\ref{fig3}(a).
Note that the relation between $-M$ and $\Delta$ is given by $-M(\phi_h)=N_0\Delta^2(\phi_h)/2H_{c2}(\phi_h)$ at $T=0$ near $H_{c2}$.
The anisotropy in $\Delta$ becomes less pronounced as $H$ decreases.

In contrast to the $H$ dependence of $\delta\Delta$, the positive $\delta n_{\rm node}$ increases as $H$ decreases.
In other words, more quasiparticles are thermally activated near nodes when the field is applied in the anti-nodal direction as compared with the case of ${\bm H}\parallel {\bm k}_{\rm node}$.
This is understood by the so-called Doppler shift argument as follows, which is valid for a low density of vortices at low temperatures\cite{Vekhter99}.
The local quasiparticle with the momentum ${\bm k}$ has the energy spectrum $E_{\bm k}+{\bm v}_{\rm s}\cdot{\bm k}$ in the supercurrent flowing around vortices with the velocity ${\bm v}_{\rm s}$.
This energy shift gives rise to a finite DOS near the gap nodes if $|{\bm k}\cdot{\bm v}_{\rm s}|>|\Delta({\bm k})|$.
Since ${\bm v}_{\rm s}\perp{\bm H}$, the number of the relevant gap nodes becomes largest in the case of ${\bm H}\perp{\bm k}_{\rm node}$.
Thus, we have $\delta n_{\rm node}>0$.
Then, the reduction of the superfluid density leads to increase of the penetration depth as well as the magnetic induction.
As a result, we have smaller $-M$ for ${\bm H}\perp{\bm k}_{\rm node}$ than that for ${\bm H}\parallel{\bm k}_{\rm node}$.

The temperature dependence of $-M$, $\delta\Delta$ and $\delta n_{\rm node}$ shows a similar behavior as shown in Fig.~\ref{fig3}(b), hence, a similar argument holds for $T$ dependence as well.
We then conclude that in the presence of the gap nodes, a sign change should occur in the amplitude of $-M$ as a function of $H$ and $T$.
The sign of the oscillatory amplitude has an important information about the direction of gap nodes.

\begin{figure}
\includegraphics[width=8cm]{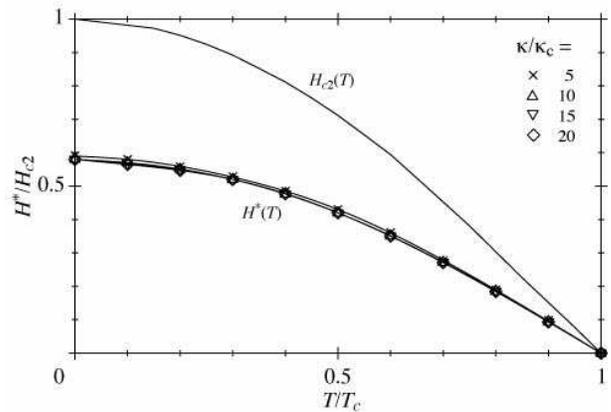}
\caption{The temperature dependence of the crossover field of reversible oscillation in the magnetization for several GL parameters.}
\label{fig4}
\end{figure}
Finally, we discuss the dependence of the GL parameter, $\kappa$.
Figure~\ref{fig4} shows the temperature dependence of the crossover field, $H^*$, for several $\kappa$'s.
The crossover field is located in the intermediate field range as a consequence of the competition described in the above.
It is almost independent of $\kappa$ at least for $\kappa/\kappa_c\ge5$.
This implies that the identification of the gap structure using the field-angle dependence of the magnetization can be applied for a wide class of the intermediate and the strong type II superconductors.

In summary, we have shown the interplay between the in-plane anisotropy of the magnetization and the nodal gap structure with four line nodes.
The field-angle dependence of the magnetization, $M$, exhibits a clear four-fold oscillation, as a direct consequence of the nodal gap structure.
The high-field oscillation in $-M$ shows a minimum when the field is applied along the nodal direction, while the low-field one has a maximum in the same field direction.
The sign change in the oscillatory amplitude can be understood by the competition between the effect of the $H_{c2}$ anisotropy and the low-field quasiparticle excitations near nodes at low temperatures.
As a result, the crossover field appears in the intermediate field range, and is almost independent of the Ginzburg-Landau parameters.
The temperature dependence shows a similar sign change.
All these features suggest that the anisotropy of the magnetization can be a good candidate to identify the gap structure directly for a wide class of type II superconductors.

\end{document}